# Specification of the Q Hypothesis:
# An Alternative Mathematical Foundation for Physics

Paul J. Werbos[1]

A "theory of everything," in modern physics, is a mathematical theory which attempts to explain or fit all of the laboratory data available today, from the level of elementary particles and nanochips, to the level of gravitational effects between galaxies. Physics today appears to have only two contenders for a theory of everything – superstring or n-brane theory, and a hoped-for merger of quantum loop gravity with the "standard model of physics" (QCD+EWT). This brief paper will attempt to specify a third candidate, which I call "the Q hypothesis."

This paper will not argue for the *truth* of the hypothesis. It will include a few remarks to explain it (with citations to more extensive explanations) , and discuss how the hypothesis might be used in the end. In the end, my claim is that this hypothesis, like the classic Wigner hypothesis[2], may have *computational* and *empirical* value, in helping us to explore and think about the universe. Religious commitment for (or against) specific theories of everything will not really help our fundamental theoretical understanding now, any more than it did when Tycho Brahe proposed his own sacred and elegant theories of strings in the heavens. The simplicity of the hypothesis may be somewhat shocking to some at first, but there is substantial analysis behind it, and the obvious questions have been considered.

An appendix added in April, 2008, explains more about the background behind the hypothesis, and the reasons why the earlier P hypothesis currently seems preferable.

## Specification

1. The theory is defined by starting from a *base theory* and making two extensions.

**2. The base theory** is a classical field theory (CFT) in the spirit of Einstein. We postulate a set of smooth continuous fields $\varphi_i(\underline{x},t)$ over flat Minkowski space, for $i = 1$ to $N$, where $N$ is some finite number. The fields form a mathematical vector $\underline{\varphi}$. In specific variations of this, the components of $\underline{\varphi}$ will of course be grouped in ways that include relativistic vectors, some under "topological constraints" (such as the Skyrme constraint that $|\underline{v}|^2=1$ for a vector $\underline{v}$ made up of some of the components of $\underline{\varphi}$), relativistic scalars and so on. We assume that these fields are governed by the classical Lagrange-Euler equations for a specific Lagrangian density $\mathcal{L}(\underline{\varphi}, \partial_\mu \underline{\varphi})$, which may be represented equivalently in terms of a Hamiltonian density $\mathcal{H}(\underline{\varphi}, \underline{\pi})$. The state of the universe S(t) at any time t is defined as the set of values of $\underline{\varphi}(\underline{x}, t)$ and of $\underline{\pi}(\underline{x}, t)$ across all points $\underline{x}$ at time t.

---

[1] The views expressed here are those of the author, not those of his employer; however, as work produced on government time, it is in the "government public domain." This allows unlimited reproduction, subject to a legal requirement to keep the document together, including this footnote and authorship. Some related material is posted at www.werbos.com.

[2] Wigner [1] proposed a way to interpret the wave function for a single electron as a probability distribution for the position *and* momentum of the particle, as a classical point particle. Wigner's interpretation never got far as a theory of physics, but is used ever more widely as an important and powerful computational tool in quantum optics and nanophotonics.

**2a. First extension**: append a set of m stochastic sources and sinks $s_i(\underline{x}, t)$ to the Lagrange-Euler equations, with $0 \leq m \leq N$. More precisely, if our N specific Lagrange-Euler equations are properly ordered, then we simply append "+$s_i(\underline{x}, t)$" to the i-th Lagrange-Euler equation. The m sources/sinks, $s_1$ through $s_m$, form a mathematical vector $\underline{s}$. $\underline{s}(\underline{x}, t)$ is defined as a source of "continuous white noise," governed by a Gaussian distribution $N(0, \Sigma_s)$, *similar to* the usual continuous white noise sources familiar from everyday engineering and from (forwards) stochastic differential equations [21]. However, in this case, $\underline{s}(\underline{x}, t)$ is a *time-symmetric* source of white noise, as will be discussed in item 3.

**Remark** – the hypothesis is that a proper choice of Lagrangian and of $\Sigma_s$ is sufficient to reproduce the standard model of physics, for the range of experiments which it actually predicts well, and also to explain certain experiments which it does not explain. In addition, the claim is that the resulting theory is mathematically well-specified without any axioms related to "regularization" and "renormalization."

**2b. Second extension:** the Lagrangian density and the Lagrange-Euler equations should be "metrified" by using the exact same procedure used by John Wheeler in his "already unified field theory"[2], for which he received the Nobel Prize. The point here is that the natural unification of CFT and general relativity is already very clear, very straightforward and well-established. (From an objective viewpoint, there is no reason to assume that the Fock Space version of this unification must be so simple and clear as the Fock space version of a quasilinear theory like the standard model of physics. We do not really need a simple unification in Fock space, if the underlying axiomatic version in CFT is simple.)

**3. In making statistical predictions** which depend on unobserved, microscopic variables (like the $s_i$ of axiom (2a), or the initial values of field variables dual to what we observe/control in setting up "initial conditions" to a scattering experiment), we should *give up the ad hoc classical procedure of assuming time-forwards local Markov process dynamics*. In my view [3], this ad hoc procedure is the real reason why CFT *appears* to be inconsistent with experiments like the classic "Bell's Theorem" experiments proposed by Clauser et al, etc. In fact, when one assumes both "A" and "not A" at the same time, one can derive all kinds of untenable predictions. When the CFT itself is symmetric with respect to time reversal T (or very close to symmetric), it is grossly inconsistent to assume at the same time that microscopic flows of "causality" only run in one time direction. *One should derive the statistics from the CFT proper (and global boundary conditions in past and future both), instead of using the ad hoc, convenient but nonviable classical assumption.*

More precisely then – the calculation of statistics both for $\underline{s}$ and for macroscopic "measurement" events like passage through a polarizer should be based on local Markhov Random Field (MRF) mathematics – MRF calculations over Minkowski space – rather than local Markhov Process (MP) mathematics. Here I am proposing that we replace one statistical model over continuous space-time or over a graph of measurement events with another; however, it is easier to understand the distinction between local MRF and local MP by considering a simple example over a space-time lattice, over one discrete spatial dimension ($i_x = -\infty, \ldots, -1, 0, 1, \ldots +\infty$) and one discrete time dimension ($t = -\infty, \ldots, -1, 0, 1, \ldots +\infty$). In the simplest local MP, the probability of a state $\underline{\varphi}(i_x, t)$ is given by:

$$\Pr(\underline{\varphi}(i_x,t)) = f(\underline{\varphi}(i_x-1,t-1), \underline{\varphi}(i_x,t-1), \underline{\varphi}(i_x+1,t-1)) \qquad (1)$$

But in the simplest local MRF over space time, it is given by:

$$\Pr(\underline{\varphi}(i_x,t)) = f(\underline{\varphi}(i_x-1,t), \underline{\varphi}(i_x,t-1), \underline{\varphi}(i_x,t+1), \underline{\varphi}(i_x+1,t)) \qquad (2)$$

Early work on "stochastic quantization[3]" showed how continuous space versions of this concept can obey the obvious requirements for finiteness and relativistic invariance and so on.[4]. See [3] for discussions of how this maps into quantum measurement experiments.

**4. Most of the major predictions** of the standard model (EWT+QCD) can generally be mapped into predictions of spectra and of "scattering states[5]" – statistical equilibria which bypass the issue of quantum measurement. If we assume that the rules of quantum measurement *should be derived from (quantum or CFT) dynamics and from boundary conditions in any case*, rather than ad hoc assumptions, the main gap here is to prove that the Q hypothesis can replicate (or improve upon) these kinds of predictions.

**Remark** – Traditional quantum electrodynamics (QED), combined with the traditional Copenhagen model of quantum measurement, *does not come even close to fitting all the empirical data available in the laboratory today.* There is a huge body of empirical data and real-world devices which can only be modeled and designed today by assuming a different practical but general theory called "cavity QED" [6]. For obvious reasons, the early developers of this theory did not stress how large and important the deviations are from the usual QED and the standard model of physics itself. The assumed Hamiltonian operator is not even "local" in the way that the usual Hamiltonian of QED is; it cannot be represented as an integral over space of a product of local field operators! Nevertheless, I hypothesize that this unpleasant situation can be resolved by using a local Hamiltonian related to the standard model (as described below) *but changing the measurement formalism* (as discussed in papers posted at [3]).

**5. Though the universe is governed by** a classical CFT (with extensions 2a and 2b), we totally lack the ability to eliminate microscopic "subquantal" fluctuations – similar to traditional thermodynamic fluctuations but acting symmetrically in time – in the states of the fields. In other words, "the microscopic universe is thermalized."

---

[3] This type of "stochastic quantization" is totally different from what goes by that name in signal processing, involving mapping signals to a kind of tessellation for purposes of data compression.

**6. For any "successful" bosonic QFT**, based on quantizing field variables $\varphi_1, \ldots, \varphi_N$, any density matrix $\rho$ corresponding to an equilibrium localized or scattering state should be interpreted to represent a statistical ensemble of possible "classical" states S defined as follows:

$$\underline{v}(S) = \exp\left( c \sum_{j=1}^{n} \int (\theta_j(\underline{p}) + i\tau_j(\underline{p})) a_j^+(\underline{p}) d^d \underline{p} \right) |0\rangle , \qquad (3)$$

where d is the number of spatial dimensions (i.e. $\underline{x} \in R^d$) and:

$$\theta_j(\underline{p}) = \sqrt{w_j(\underline{p})} \int e^{-i\underline{p}\cdot\underline{y}} \varphi_j(\underline{y}) d^d \underline{y} \qquad (4)$$

$$\tau_j(\underline{p}) = \frac{1}{\sqrt{w_j(\underline{p})}} \int e^{-i\underline{p}\cdot\underline{y}} \pi_j(\underline{y}) d^d \underline{y} \qquad (5)$$

$$w_j(\underline{p}) = \sqrt{m_j^2 + |\underline{p}|^2} \qquad (6)$$

$$\Pr(S) = <\underline{v}(S) | \rho | \underline{v}(S)> \qquad (7)$$

Equations 3 through 6 come from [7]. Equation 7 is the core of the definition of the "Q hypothesis."

**Remark 1** – In [7], I considered an alternative hypothesis, the "P hypothesis," based on mapping *from* a probability distribution Pr(S) to a density matrix $\rho$ by:

$$\rho = \int \frac{\underline{v}(S)\underline{v}^H(S)}{|\underline{v}(S)|^2} \Pr(S) d^\infty S , \qquad (8)$$

This exactly reproduced some of the key equilibrium properties of bosonic QFT, enough to satisfy key axioms used by Weinberg in his derivation of QFT [8]. In particular, we proved that:

$$\mathrm{Tr}(\rho H_n) = < H(S) > \qquad (9)$$

where $H_n$ is the normal-product form of the Hamiltonian operator, where H(S) is the total energy (Hamiltonian) of the CFT state S, where the density matrix $\rho$ is calculated by equation 8, and where the angle brackets denote the (classical) expectation value.
    Nevertheless, in later calculations [9], my partner and I discovered puzzling discrepancies between the equilibrium predictions of QFT and those of the P hypothesis. These were essentially the same as the "quantum correction terms" for the CFT versus QFT mass of model solitons, as described by Coleman [10] and by Rajaraman [11].

The paradox may be explained as follows: when we look for states ρ of *minimum energy* in QFT, we are allowed to consider states ρ which cannot be reached by *any* allowable (nonnegative) probability distribution Pr(S); for *reachable* states ρ, the classical and the quantum energy predictions are the same, but not all states are reachable.

The Q hypothesis eliminates this problem, because equation 7 always yields an acceptable Pr(S) for *any* density matrix ρ allowed in QFT. However, it assumes that we cannot reach *all possible mathematically well-defined states S(t)* in actual experiments. That is why assumption (5) is an essential part of the hypothesis.

**Remark 2** – Some theoretical physicists may find equation 9 to be quite astounding. But several years after we derived it – and discussed it with many, many others – we found out that it is basically just a generalization of well-established results for the "P" mapping developed by Glauber many years before for use in modeling electromagnetism (light). Glauber's "P" and "Q" mappings were a major part of the work which won him the Nobel Prize in 2005. They are a major staple of modern quantum optics [12,13,14]. I first considered using equation 7 by considering how the usual Q mapping can also be generalized, and used to overcome the discrepancies of the prior work.

From the work in quantum optics [12,13,14], it is well-known that the Q probability distribution (pdf) is a "fuzzified" version of the corresponding P distribution. More precisely, the Q mapping allows us to reach any statistical mixture of classical pdfs defined by:

$$\Pr(S) = c \, \exp\left( \int \left( -\left| \underline{\varphi}(\underline{p},t) - \underline{\varphi}_0(\underline{p},t) \right|^2 - \left| \underline{\pi}(\underline{p},t) - \underline{\pi}_0(\underline{p},t) \right|^2 \right) d^3\underline{p} \right) \qquad (10)$$

for a set of base field values {$\underline{\varphi}_0(\underline{x})$, $\underline{\pi}_0(\underline{x})$} at time t, representing a base state $S_0$. Again, we may reach any *mixture* of pdfs like equation 7, across a set of possible base states $S_0$, but in nondegenerate cases we would expect energy to reach a minimum for classical pdf with a definite base state $S_0$.

**Remark 3** – In general terms, I hypothesize that equation 10 is the result of "thermalization of the universe." In fact, it is well known that a Boltzmann distribution about a local minimum of energy can be well-approximated by a Gaussian distribution, in the local neighborhood. The $w_j$ factors in equations 4 and 5, and the integration over **p**, eliminate problematic cross-correlation effects. Nevertheless, this only works if the units used to describe each specific type of fundamental soliton are scaled to give a unit variance in equation 10; this suggests that the apparent multiplicity of fields in QFT might be explained in part as the result of different scaling of different solitons based on a smaller number of underlying fields.

**Remark 4 –** The dynamic predictions of the Q hypothesis would not be identical to those of the corresponding QFT under all circumstances. For example, the Q hypothesis would predict that zero degrees Kelvin (as presently understood) is not truly a state of perfectly zero motion. This very strong deviation from standard QED has in fact been verified empirically, in very extensive research replicated by many leading laboratories [15,16].

**Remark 5** – If we assume that the thermalization is due to boundary condition effects at (infinite) space and time, we would presumably end up with a traditional Boltzmann distribution, which contains a temperature *vector* – a vector of coefficients of H and **P** in the grand canonical ensemble. That vector provides a certain kind of preferred direction or arrow of time, violating the spirit of special relativity to some degree. But if we assume that it results from assumption (2a) above, the problem disappears. The effect is as if each "heavy point particle" is perturbed *relative to its own rest frame*. Nevertheless, no one on earth has ever measured the variation in the level of zero-temperature decoherence for systems moving near the speed of light. Thus we do not have a strong empirical basis as yet for preferring one variation over the other.

**7. Under the generalized P mapping**, the classical energy functional H maps into $H_n$, the normal-product form of the Hamiltonian. Thus the P hypothesis would assert that the CFT governing the universe have a Hamiltonian H such that $H_n$ appears to be a valid bosonic QFT. *But the mapping is different for the Q mapping*, as is well-known in quantum optics. In essence, we use the anti-normal product. Thus the classical-quantum equivalence maps between a classical energy density which is mathematically well-defined, in a clean way, and a QFT which is well-defined *only with* the addition of a kind of regularization procedure for dealing with the "zero point energy" terms which result. (In fact, such a procedure is also used in the first stage of traditional canonical quantization or in Feynman path quantization. The Q hypothesis appears more consistent with the Hawkings theory of gravity than the P hypothesis, but the theoretical and empirical issues related to mid-sized black holes, for example, are very far from being well-established as yet.) The classical *axioms* are clean, but what they map to is the fully messy reality of bosonic QFT – *with a basis for deriving and truly proving rather than assuming the regularization of zero-point terms*.

**8. The hypotheses above (parts of the Q hypothesis) make sense only if** we can replicate (or improve upon) the predictions of the standard model (EQT+QED), by using a purely bosonic field theory. That once seemed impossible, because the standard model includes fermionic fields as well as bosonic fields, and the "spin-statistics theorem" [17] seemed to rule out constructing fermionic fields from any kind of aggregation or behavior of bosonic fields. However, a large body of work (e.g. [10, 18, 19]) has shown that this is not true. For the Q hypothesis, I make the following additional subhypothesis: that we can replicate the demonstrated predictions of the standard model as the *limit* as r→0 of a *parameterized family* of bosonic (and classical) field theories defined by a Hamiltonian density $\mathcal{H}(\varphi, \pi, r)$, where each field theory for r>0 is mathematically well-defined and nonsingular even without any regularization or renormalization procedure. In effect, the passage to a limit in r could be considered as a kind of *constructive physical regularization*; if we assume a very small nonzero value of r, we arrive at a theory which fits the empirical data but is well-defined even without regularization axioms.
The Hamiltonian would be chosen so as to yield "solitons" whose radius is roughly proportional to r, but whose other properties do not change substantially as r changes in the neighborhood of zero.

**Remark** – Why is renormalization and regularization unavoidable in the standard model of physics? A major reason is that a charged point particle has an infinite energy of self-repulsion. Traditional QED has no explanation for *why* nature somehow converts this to a finite physical mass-energy. Where does the deus-ex-machina infinite negative energy of renormalization come from? Modeling the electron (and quark) as a particle with *nonzero* radius – a "soliton" – is the obvious and natural solution. A key hidden reason why superstring theory overcomes this problem is that it assumes the electron has a very small but nonzero radius – but we don't need to postulate lots of unobserved hidden dimensions of space in order to obtain this benefit.

**9. In 8, it is *not assumed* or *required*** that these bosonic field theories be "superrenormalizable" or even "renormalizable" in the usual sense. To be mathematically well-defined, it is good enough that that the CFT themselves be well-posed in a reasonable sense as partial differential equations. Leaders in axiomatic QFT like Arai have long recognized that *nonperturturbative* methods will be required, in order to achieve mathematically well-defined field theories powerful enough to reflect what the standard model can predict. The underlying problem is that conventional renormalization and perturbation is based on Taylor series expansion about zero (the vacuum state) – but those kinds of Taylor series simply don't work in describing many important field systems and states, like solitons. Rajaraman [11] has stressed that a different kind of polynomial expansion – a "WKB" expansion, an expansion about a nonzero state like a soliton state – is essential to the mathematics of this class of QFT.

## Some Possible Ways of Pursuing the Q Hypothesis

As with the P hypothesis [3], the Q hypothesis has empirical implications for quantum measurement which are well-worth pursuing in their own right – particularly for areas like quantum optics. There are obvious interesting issues in basic mathematics as well.

Perhaps the most exciting possibility is that the computational methods which have been crucial to the power of modern QED engineering (photonics, chips, etc.) could be applied to the realm of strong nuclear forces. Even if the P and Q hypotheses turn out to be wrong, in the end, they do provide a kind of computable upper bound and lower bound to the energy predictions of bosonic QFTs. Furthermore, recent progress with atom and hadron lasers suggests that the same coherence effects which have radically shaped what we can do with electromagnetism might also allow us to do things with strong nuclear forces far beyond what today's megacollider two-body thinking allows us to imagine; to make this possible, the same type of mathematics needed to understand and exploit coherence effects in quantum optics may be essential.

However, where we can we get a more specific Lagrangian for a bosonic family of models that could replicate the predictions of QCD?

There are many possibilities here. (We should be happy that there is more than one possibility allowed by this framework. It is better that *empirical* data make the choice between possibilities, and that we not be restricted to only one.) The most obvious possibility is to start from the Hasenfratz/'tHooft model [18] of a fermionic soliton which emerges from the totally bosonic Lagrangian they specify. But that soliton is a "dyon" – an object of mixed electrical and magnetic charge! The most obvious next step is to try to establish the validity of Julian Schwinger's proposed extension/modification of the quark model [20], which uses these dyons in place of conventional quarks. In this effort, it is essential to incorporate Schwinger's suggestion (at the end of [20]) that gluons may also have magnetic charge. With a that modification, Schwinger pointed out that his model can actually address empirical results which the QCD has been unable to explain to this day.

One important technicality here is that bosons still give us a choice. They may be bound states of solitons in some cases, or simple radiative fields without mass in others. There is no reason why "mass-like" $m|\phi|^2$ terms cannot appear in the Lagrangians of such fields. Thus there are a variety of parameters to explore empirically. Also, it is not so clear that we really need "color" here to explain everything observed so far; Schwinger's model provides an alternative starting point whose predictions can be explored much further.

It is possible, however, that some of the key experiments here might be performed more safely in earth orbit than on the surface of the earth, until we have a better understanding of how they work (and how they interact with gravity). Fortunately, many-body experiments performed in a vacuum may not require the huge masses that we are accustomed to from earth-based supercolliders.

23. P. Werbos, *Schwinger's magnetic model of matter – can it help us with grand unification*, under submission, revised January 2008, arXiv:0707.2520

## Appendix: P Hypothesis Versus Q Hypothesis

This paper essentially just specified a new possible formulation of physics, without explaining the prior background, or whether we should actually believe it in the end. The background is provided in [22], which argues that: (1) the classic Copenhagen version of quantum field theory which most of us were taught is massively disconfirmed by empirical evidence; (2) because of certain weaknesses in the derivation of the Everett/Wheeler version, there are strong logical reasons to prefer the Backwards Time Physics (BTP) version of physics, and to follow up on empirical opportunities to discriminate between BTP and conventional Everett/Wheeler. BTP leads directly to an explanation of the modern "Bell's Theorem" experiment, which then gives us a choice: (1) we may do physics based on MW/BTP, a variation of many worlds theory; or (2) it becomes theoretically possible to go back to the program of Einstein, to represent the laws of physics as partial differential equations (PDE) *within the context of BTP*. By Occam's Razor, we should prefer the second alternative, if we can make it work, but how can we find the specific PDE which fit the massive amounts of empirical data available to us today? How can we find an empirical strategy to decide which of the two is true?

We started from the simple idea that reality might be governed by partial differential equations over ordinary Minkowski space. By reinventing and generalizing the "P" transformations well-known in quantum optics, we showed how bosonic quantum field theories in general have PDE equivalents, which in some sense are exactly equivalent [7]. (I owe great thanks to Brandt of the Army Research Labs and to Jon Dowling for making me aware of [12] and [13], two of the three modern references on the P and Q transforms, which go back to the classic work by Glauber.) However, in analyzing the situation further, we found certain discrepancy terms [9] which would affect the crucial empirical predictions for spectra and for scattering states (the S matrix).

The essence of the Q hypothesis is quite simple. We can make the discrepancy terms go away, and match any bosonic field theory (in a large class which includes anything on the table today) with a "classical" theory, *so long as we add white noise*. Thus instead of modeling the universe by PDE as such, we model the universe as stochastic PDE, with white noise terms added in a time-symmetric way, as in [21]. This is a rather neat idea, because, among other things, it makes it clear that this is not your grandmother's classical field theory. It is rigorously well-defined (given the PDE and the white noise model), but it include "God throwing dice" in a simple but nonclassical way. It is an example of what I would call "stochastic realism."

But do we really need this extra complexity, these white noise terms, in order to match the empirical evidence which physics has accumulated through the years? That is an empirical question. We would need empirical evidence in any case to estimate the size of the white noise effects. The size might well be zero, bringing us back to the original P hypothesis. That is why this paper was clear that it was formulating the Q hypothesis only as a *possibility worth exploring*.

In 2006, I did have a feeling that the Q hypothesis would probably turn out to be true. The discrepancy terms do go away in some bosonic field theories, such as electromagnetism itself; however, to model all of physics, one would need to introduce strong nonlinearities, and it appeared that strong and measurable discrepancy terms would probably arise.

But now, in April 2008, that no longer appears to be the case. In [28], I have looked more closely at the issue of how we can explain the range of phenomena explained by the standard model of physics – *and additional empirical phenomena it has problems with or cannot address at all* – by using bosonic quantum field theories which induce solitons. Since Coleman's classic 1977 paper on the equivalence of the sine-Gordon (bosonic) field theory and the Massive Thirring Model, it is now widely understood that fermionic bound states can indeed emerge in bosonic field theories; the field of "bosonization" has become an enormous though scattered enterprise.

Following up on the appendix to [28], it appears that the underlying mechanism of bosonization is really far simpler than one might imagine. One may "regularize" the usual kind of quantum field theory by mapping each fermion (at a point **x** or at a point **p**, normal or virtual) into a soliton of radius r, and then taking the limit as r goes to zero. This leads very easily to the correct energy value, in any possible state, in the limit as r goes to zero, so long as the soliton approaches the right limit as a function of r. In essence, it needs only to have the right overall mass and spin and charges and location. The only place for discrepancy terms to cause a problem is if the bosonic part of the original field theory (like quantum electrodynamics, QED) has discrepancy terms; however, for QED it is well-known that they do not. Since QED is really the only case where we have a lot of twelve-digit precision in empirical results, there is currently no empirical reason to believe that the noise terms are nonzero.

Note that this new, simple view of bosonization (for models like QED) does not really rely on complex group theoretic arguments or even the full power of antisymmetric statistics. It relies on the Pauli exclusion principle, and it relies on a combination of continuity and antisymmetric statistics, to ensure that we don't have to worry about the errors which could result if we piled two solitons on top of each other, *in the limit* as r goes to zero. Formally, when I say that "r goes to zero," I mean that the parameters of the field theory follow a path in which the radius of the soliton goes to zero, as described in [23]. Conventional relations between spin and statistics and angular momentum are implied by all of this, but are not required as axioms or assumptions.

Aside from cleaning up the details here, this leaves one major theoretical task in the program laid out in [23]: to verify what the actual properties are of selected soliton-generating theories, both in the usual quantum (many worlds) version and in the PDE versions. *Either* version provides a way of getting rid of the need for renormalization in formulating the underlying laws of physics! The two versions may well disagree, for a given Lagrangian and a given set of parameters. If the disagreement is fundamental (which currently appears unlikely), it may even provide an empirical basis for deciding between the two theories, based on which is better at fitting the nuclear mass systematics studied by MacGregor, Palazzi, and others. Nuclear mass systematics is a major empirical challenge which the current standard model cannot even address [23].

In summary, it seems that we have finally reached a point where we can go back to nature – to empirical phenomena – to discriminate between different fundamental theories of physics, and where simpler theories seem likely to fit as well or better than the elaborate imaginative creations of recent years.